\documentclass[useAMS,usenatbib]{mn2e}

\usepackage{epsfig}
\usepackage{amsmath}

\newcommand{\for}[1]{\begin{equation} #1 \end{equation}}

\def\gsim{\mathrel{\raise0.35ex\hbox{$\scriptstyle >$}\kern-0.6em\lower0.40ex\hbox{{$\scriptstyle \sim$}}}} 
\def\lsim{\mathrel{\raise0.35ex\hbox{$\scriptstyle <$}\kern-0.6em\lower0.40ex\hbox{{$\scriptstyle \sim$}}}}

\title[High resolution 870~$\mu$m source counts from the ALMA/LESS survey]{
An ALMA survey of submillimetre galaxies in the Extended Chandra Deep Field South: 
High resolution 870~$\mu$m source counts}

\author[A.\ Karim et al.]
{\parbox[h]{\textwidth}
{A.\ Karim,$^{\! 1}$\thanks{E-mail: alexander.karim@durham.ac.uk (AK)}
A.\,M.\ Swinbank,$^{\! 1}$ 
J.\,A.\ Hodge,$^{\! 2}$
Ian Smail,$^{\! 1}$ 
F.\ Walter,$^{\! 2}$
A.\,D.\ Biggs,$^{\! 3}$ 
J.\,M.\ Simpson,$^{\! 1}$
A.\,L.\,R.\ Danielson,$^{\! 1}$
D.\,M.\ Alexander,$^{\! 1}$
F.\ Bertoldi,$^{\! 4}$ 
C.\, de Breuck,$^{\! 3}$
S.\,C.\ Chapman,$^{\! 5}$
K.\,E.\,K.\ Coppin,$^{\! 6}$ 
H.\ Dannerbauer,$^{\! 7}$
A.\,C.\ Edge,$^{\,1}$
T.\,R.\ Greve,$^{\! 8}$
R.\,J.\ Ivison,$^{\! 9,10}$
K.\, K.\ Knudsen,$^{\! 11}$ 
K.\,M.\ Menten,$^{\! 12}$ 
E.\ Schinnerer,$^{\! 2}$ 
J.\,L.\ Wardlow,$^{\! 13}$
A.\ Wei\ss,$^{\! 12}$ 
P.\ van der Werf$^{14}$ 
}
\vspace*{6pt} \\ 
$^1$Institute for Computational Cosmology, Durham University, South Road, Durham, DH1 3LE, UK\\
$^2$Max-Planck-Institut f\"ur Astronomie, K\"onigstuhl 17, D-69117 Heidelberg, Germany\\
$^3$European Southern Observatory, Karl-Schwarzschild Strasse 2, D-85748 Garching, Germany\\
$^4$Argelander-Institute of Astronomy, Bonn University, Auf dem Huegel 71, D-53121 Bonn, Germany \\
$^5$Institute of Astronomy, University of Cambridge, Madingley Road, Cambridge CB3 0HA\\
$^6$Department of Physics, McGill University, 3600 Rue University, Montréal, QC H3A 2T8, Canada\\
$^7$Universit\"at Wien, Institut f\"ur Astrophysik,  T\"urkenschanzstra\ss e 17, 1180 Wien, Austria\\
$^{8}$University College London, Department of Physics \& Astronomy, Gower Street, London, WC1E 6BT, UK \\
$^9$UK Astronomy Technology Centre, Science and Technology Facilities Council, Royal Observatory, Blackford Hill, Edinburgh EH9 3HJ\\
$^{10}$Institute for Astronomy, University of Edinburgh, Blackford Hill, Edinburgh EH9 3HJ\\
$^{11}$Department of Earth and Space Sciences, Chalmers University of Technology, Onsala Space Observatory, SE-43992 Onsala, Sweden\\
$^{12}$Max-Planck Institut f\"ur Radioastronomie, Auf dem H\"ugel 69, D-53121 Bonn, Germany\\
$^{13}$Department of Physics \& Astronomy, University of California, Irvine, CA 92697, USA\\
$^{14}$Leiden Observatory, Leiden University, PO Box 9513, 2300 RA Leiden, Netherlands\\
}

\begin{document}

\date{Accepted 2013 January 31.  Received 2013 January 30; in original form 2012 September 30}

\pagerange{\pageref{firstpage}--\pageref{lastpage}} \pubyear{2013}

\maketitle

\label{firstpage}

\begin{abstract}
We report the first counts of faint submillimetre galaxies (SMG) in the 870-$\mu$m band derived from arcsecond resolution observations with the Atacama Large Millimeter Array (ALMA). We have used ALMA to map  a sample of 122 870-$\mu$m-selected submillimetre sources drawn from the $0.5^\circ \times 0.5^\circ$ LABOCA Extended {\it{Chandra}} Deep Field South Submillimetre Survey (LESS). These ALMA maps have an average depth of $\sigma_{\rm 870\mu m}\sim 0.4$\,mJy, some $\sim 3\times$ deeper than the original LABOCA survey and critically the angular resolution is more than an order of magnitude higher, FWHM of $\sim$\,1.5$''$ compared to $\sim$\,19$''$ for the LABOCA discovery map. This combination of sensitivity and resolution allows us to precisely pin-point the SMGs contributing to the submillimetre sources from the LABOCA map, free from the effects of confusion. We show that our ALMA-derived SMG counts  broadly agree with the submillimetre source counts  from previous, lower-resolution single-dish surveys, demonstrating that the bulk of the submillimetre sources are not caused by blending of unresolved SMGs. The difficulty which well-constrained theoretical models have in reproducing the high-surface densities of SMGs, thus remains.   However, our observations do show that all of the very brightest sources in the LESS sample, $S_{\rm 870\mu m}\gsim $\,12\,mJy, comprise emission from multiple, fainter SMGs, each with 870-$\mu$m fluxes of $\lsim $\,9\,mJy . This implies a natural limit  to the star-formation rate in SMGs of 
$\lsim $\,10$^3$\,M$_\odot$\,yr$^{-1}$, which in turn suggests that the space densities of $z>$\,1 galaxies with  gas masses in excess of $\sim $\,$5 \times 10^{10}$\,M$_\odot$ is $< 10^{-5}$\,Mpc$^{-3}$. We also discuss the influence of this blending on the identification and characterisation of the SMG counterparts to these bright submillimetre sources and suggest that it may be responsible for previous claims that they lie at higher redshifts than fainter SMGs.
\end{abstract}

\begin{keywords}
galaxies: starburst, galaxies: evolution, galaxies: high-redshift
\end{keywords}

\section{Introduction}

The first deep surveys for extragalactic submillimetre sources \citep{SMAI97, HUGH98, BARG98} uncovered high number densities of submillimetre sources at mJy-flux limits and subsequent spectroscopy determined a median redshift of $z \sim 2.5$ for the radio-detected subset of the submillimetre galaxy (SMG) population \citep{CHAP05}.  At these high redshifts, the submillimetre fluxes of these sources correspond to far-infrared luminosities of $>$\,10$^{12-13}$\,L$_\odot$, placing them in the ultra-luminous or hyper-luminous infrared galaxy (ULIRG, HLIRG) classes.   These joint 850-$\mu$m and radio-selected samples remain the best-studied SMGs and it has been claimed that they host up to half of the star formation occurring at $z \gsim $\,2 \citep[e.g.,][]{HUGH98, BLAI99, CHAP05} and may be linked to QSO activity and the formation of massive galaxies at high redshift \citep[e.g.,][]{SWIN06, ALEX08, HICK12}. If true then SMGs are an essential element in models of galaxy formation. In fact the first theoretical attempts to reproduce basic properties of SMGs, in particular the 850-$\mu$m number counts, required radical alteration of the prescription for starbursts in well-constrained galaxy formation models \citep[e.g.,][]{BAUG05, GRAN06}, demonstrating the potential power of SMGs as a constraint on galaxy evolution theories. 

One concern about the use of the  850-$\mu$m number counts as a fundamental constraint on galaxy formation models is that these are derived from  low spatial resolution (typically $\sim $\,15--20$''$ full width at half maximum, FWHM), single-dish surveys.  This low resolution  means that  it is possible that several faint sources within a beam will appear as a single brighter source,  changing the shape of the number counts, most critically by potentially producing a false tail of bright sources. A number of attempts have therefore been made to obtain high angular resolution continuum imaging through interferometric observations of individual submillimetre sources \citep[e.g.,][]{GEAR00, LUTZ01, DANN02, YOUN08A, YOUN08B, WANG11} and (nearly) flux-limited samples \citep[][]{YOUN07, YOUN09,SMOL12B,BARG12}. These  observations have indeed shown that a number of bright submillimetre sources actually comprise emission from multiple SMGs. However, the conclusions from many of these studies have been weakened by a number of factors. Firstly, both the modest numbers of sources studied and the fact that the discovery surveys underlying these studies are typically shallow or restricted to small areas, has meant it has not been possible to conclusively test the shape of the bright-end of the submillimetre source counts. Secondly, many of the follow-up observations of 870\,$\mu$m-selected submillimetre sources have been carried out at longer wavelengths (typically $>$\,1.2\,mm). This has led to ambiguous results, especially when comparing the single-dish and interferometer-based fluxes for sources, and so limits the conclusions that can be drawn about their multiplicity.   However, one particularly noteworthy study is that recently published by \citet{BARG12} (see also \citealt{WANG11}) which used the Submillimeter Array (SMA) at 850\,$\mu$m to observe sixteen 850\,$\mu$m-selected submillimetre sources with fluxes $>3\,$mJy ($>4\,\sigma$) from a SCUBA survey of 110 arcmin$^2$ within the Great Observatories Origins Deep Survey North (GOODS-N; \citealp{GIAV04,WANG04}). This wavelength-matched study, yielding interferometric resolution of $\gsim 2''$, showed the best evidence yet  for an increased incidence of multiple SMGs in submillimetre sources at bright 850-$\mu$m fluxes. However, better statistics are needed given the small sample and the substantially larger beam of the SMA compared to that of the bolometer might give rise to serendipitous detections not associated  with the underlying submillimeter source.

The issue of reliable SMG counts, needed to robustly constrain the theoretical models, is obviously an area where
the Atacama Large Millimeter Array (ALMA) will have significant impact.  In 2004 we therefore started planning  a survey to provide a large, flux-limited sample of submillimetre sources over a wide area in a field with excellent visibility from ALMA.  The field chosen was the 0.5$^{\circ}$\,$\times$\,0.5$^{\circ}$  Extended {\it Chandra} Deep Field South  (ECDFS) which has the most extensive multi-wavelength coverage of any large-area extragalactic region in the southern hemisphere.  The result was the LABOCA ECDFS submillimetre survey \citet{WEIS09} (LESS and W09 hereafter), which obtained a deep, $\sigma_{\rm 870\mu m}\sim $\,1.2\,mJy, homogeneous 870-$\mu$m  map of the full ECDFS detecting 126 submillimetre sources.  

As the next step, in ALMA Cycle 0 we observed 122 of the 126 submillimetre sources from LESS using ALMA in its compact configuration.  These data and the resulting SMG catalogue are presented in \citet[subm.; H13 hereafter]{HODG13}. Critically, these observations were carried out at the same wavelength as the LABOCA survey. The resulting ALMA maps yield unambiguous identifications for a large fraction of the submillimetre sources, directly pin-pointing the SMG(s) responsible for the 870-$\mu$m emission to within $<$\,0.3$''$ \citep[e.g.\ ][]{SWIN12}. The spatial resolution achieved by our observations ($\sim$\,1.5$''$ FWHM) corresponds to an order of magnitude improvement over the single-dish LABOCA survey. It thus provides an ideal data set to determine the influence of multiplicity on  the form of the 870-$\mu$m SMG counts, as required to enable a  reliable comparison to model predictions \citep[e.g.,][]{BAUG05}.

In this paper we analyze these ALMA maps to derive number counts for SMGs and compare these to both previous source counts from single-dish submillimetre surveys and to predictions from  theoretical models.  We adopt a cosmology with $\Omega_{\Lambda} =$\,0.73, $\Omega_{M}=$\,0.27, and H$_{0}=$\,72\,km\,s$^{-1}$\,Mpc$^{-1}$ in which a scale of 1$''$ corresponds to a physical separation of $\sim $\,8.4\,kpc at a redshift of $z=$\,2.

%
%
\begin{figure*}
  \centerline{
    \psfig{file=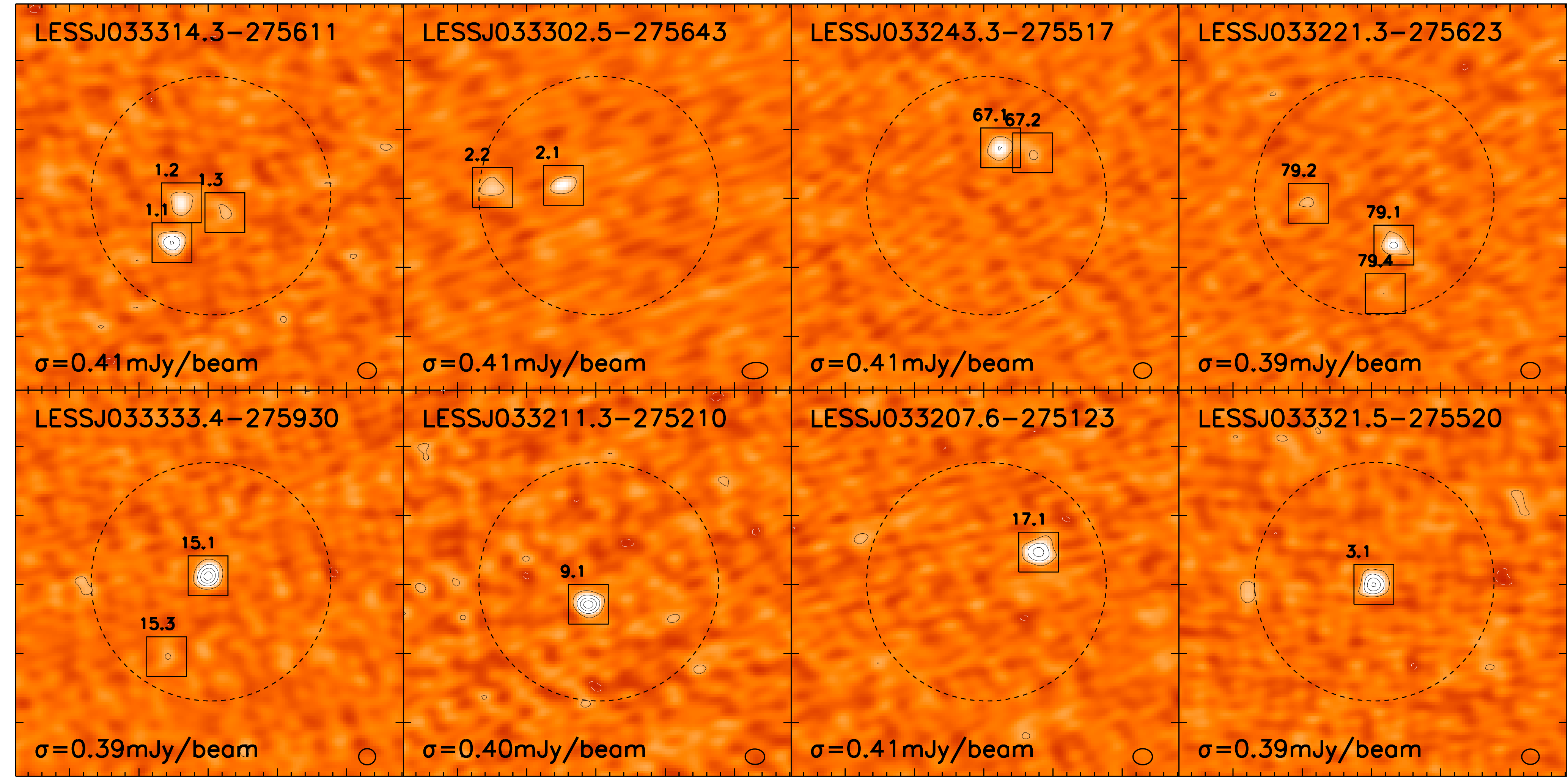,angle=0,width=\textwidth}
    }
\caption{Examples of the 870-$\mu$m ALMA continuum maps towards eight of the submillimetre sources from the LESS survey. In each map we identify all of the sources with S\,/\,N\,$>$\,3.5\,$\sigma$ (squares labeled by their catalog number; see H13). The ALMA data unambiguously locates the SMGs to a precision of $<$\,0.3$''$ and to flux limits of $\sim$\,2\,mJy\,beam$^{-1}$ ($\sim $\,3.5--7\,$\sigma$). The upper row shows a selection of maps containing multiple detections, including the maps towards the two brightest LESS sources in our sample (see text for details).
Different SMGs found in a given map are typically separated by $>6''$, corresponding to a minimum separation of $\gg 40''$. The third panel shows the very closest projected distance found in this survey ($2.6''$), corresponding to at least $\sim 20$\,kpc of separation if both sources reside at the same redshift.
The lower row shows those maps that contain the individually brightest ALMA SMGs, determining the bright end of our source counts (see \S \ref{sec:resdis}). Note that these SMGs are not necessarily associated with the brightest LESS sources. Positive (negative) contours on each map are shown in black (white) and start at $(-)3\sigma$ and are incremented by $(-)5\sigma$. The 1-$\sigma$ noise in the map is shown in the bottom left corner of each panel. Each map is $25.6''$ across and we show the primary beam (dotted circle) encompassing the radius at which the ALMA antenna sensitivity drops to 50\%, closely resembling the LABOCA beam.}
\label{fig:maps}
\end{figure*}

\section[]{Observations and  calibration}

Of the 126 submillimetre sources detected in the LESS survey, 122 were observed during Cycle 0 using the  band 7 receivers in ALMA's compact array configuration.  The available 8-GHz bandwidth was centered at an observed frequency of 344\,GHz (i.e.\ 870\,$\mu$m) and we employed a dual polarization setup. This choice of observing frequency enables a direct comparison to the flux densities of the submillimetre sources measured in the original LABOCA  survey. The observing campaign was carried out in eight observing blocks (ALMA measurement sets; MS hereafter) between 2011 October 18 and 2011 November 3. Typically 15 antennas were available for each block.

The quasar B\,0402$-$362 (J0403$-$360) was used for phase calibration and Mars, as well as Uranus, have been used to calibrate the absolute flux scale. Bandpass calibration was generally performed using observations of B\,0537$-$441 (J0538$-$440). Each science field, centered on the catalogued position of a given LESS source from W09, was observed for a total of $\sim $\,120\,seconds. The data were processed with the Common Astronomy Software Application ({\sc{casa}}; \citealt{MCMU07}) and imaged using the {\sc{clean}} algorithm within {\sc{casa}}. A detailed description of the raw data and its calibration as well as imaging is presented in H13. 

The field of view -- defined as the FWHM of the ALMA antenna reception pattern around the phase center and referred to as primary beam in the following -- is $17.3''$ in diameter. \footnote{Accordingly, the flux density at a given position in the resulting map can be corrected by multiplication with the corresponding factor derived from an inverse $17.3''$ FWHM Gaussian. We will note in the following when primary-beam corrected fluxes are used. Also note that the FWHM used, based on actual beam measurements is slightly smaller than theoretically expected for a 12-m antenna (in absence of other publicly available information on this matter we refer to ALMA help desk ticket CSV-1014 for further information).}
Each  map has a pixel size of 0.2$''$ and a total extent of 128 pixels in each dimension, sufficient to cover the primary beam and encompass the error-circles of the submillimetre sources from the LESS maps, $\lsim $\,5$''$ (W09), even in confused situations. The average  root mean square (rms, $\sigma$) of the background noise in the maps is $\sigma \sim $\,0.4\,mJy\,beam$^{-1}$ -- a factor $\sim $\,3 deeper than the original LABOCA observation.  Using natural weighting we achieve a typical restoring (clean) beam of $\sim$\,1.8$''\times$\,1.2$''$, although a small number of low elevation ($\ll $\,30\,deg) observations lead to much larger beam ellipticities and the corresponding  image products are typically much noisier than our median maps, producing  a tail in the noise distribution extending beyond 0.6\,mJy\,beam$^{-1}$. In the following -- unless explicitly stated otherwise -- we will focus on the sub-set of 88 ``best'' maps selected from two important, but not mutually exclusive, selection criteria: beam-axial ratio $< $\,2 and rms noise level $\sigma< $\,0.6\,mJy\,beam$^{-1}$.\footnote{Note that we analyze those maps that do not comply with our selection criteria in the same way as the 88 ``best'' maps, including the identification of sources as described in the following. Bright sources detected at sufficient significance in those maps form part of a supplementary ALMA catalogue which is not used in this paper unless explicitly stated. See H13 for further details.} The distribution of targets between and within each MS was chosen so that problems with the observations of any particular MS would not bias our sample and hence our ``best'' sample represents  a random sampling of the LESS catalogue, yielding an unbiased view of the properties of submillimetre sources as a function of 870-$\mu$m flux.  Figure~\ref{fig:maps} shows  examples of the calibrated and cleaned maps used in this study. The full source catalogue and all maps are presented in H13.

\section[]{Identification of ALMA sources}

\subsection[]{Source extraction and characterization}
In order to detect SMGs in our calibrated and cleaned maps we use an automated scheme (described in detail in H13). Our {\textsc{idl}}-implemented source extraction software first identifies individual signal peaks above a 2.5-$\sigma$ threshold which are used as the basis to model the emission in that region using a Metropolis-Hastings Markov chain Monte Carlo (MH-MCMC) algorithm to determine the best six parameter fit for an elliptical Gaussian\footnote{These six parameters are peak flux density, pixel position of the peak, minor axis extent, axial ratio and position angle} to describe the underlying flux distribution within a 2$'' \times $\,2$''$ region. This is large enough to recover any extended sources but small enough to resolve double sources. Whilst we attempt a full six parameter fit, at the resolution of our survey ($\sim $\,$1.5''$ corresponding to physical scales of $\sim $\,12\,kpc at $z>$\,1), many sources are unlikely to be resolved. We therefore repeat the fitting process using a simple elliptical point-source model with only three free parameters and source extents as well as orientation fixed to the synthesized beam parameters. We find that the peak flux densities from the simple fit generally agree with the integrated flux densities given their full fitting errors, we find a $>$\,2$\sigma$ integrated flux excess in just one source, suggesting it may be resolved. For all other sources we therefore adopt the peak flux density from the three parameter model as our best estimate of the source flux.   

For each parameter in the fit its posterior distribution determines the fitting error, after correcting for auto-correlation in the Markov chain. These  uncertainties are taken into account when determining the full measurement error for the integrated source flux density. The recipes we follow to determine the full uncertainty for elliptical source fits in the presence of correlated noise in radio maps have been motivated by \citet{COND97} (see also \citealt{WIND84}; \citealt{HOPK03}, \citealt{SCHI04,SCHI10}; \citealt{KARI11}).

\subsection{Source extraction efficiency and flux recovery}
\label{sec:sim}

%
%
\begin{figure}
  \centerline{
  \includegraphics[trim=0.0cm 0cm 0.0cm 1.0cm, clip, angle=0,width=0.55\textwidth]{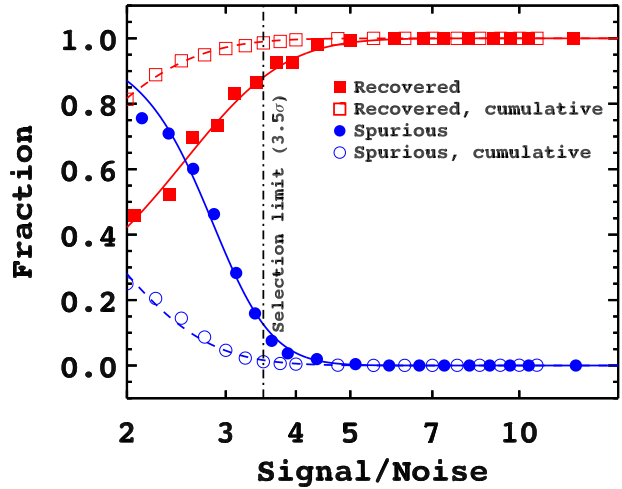}
    }
\caption{Results from artificial source simulations to test the reliability of our source extraction procedure. After removing  $>$\,2.5$\sigma$ sources identified by our automatic source finder we insert in each map five  well-separated sources at random positions within the primary beam, then run our source extraction code and repeat this process 16 times.  
Red squares denote the fraction of inserted sources that are recovered within a given narrow signal-to-noise ratio (SNR) bin while blue circles indicate the fraction of extracted sources which are unassociated within 0.8$''$ with an inserted source. Above the 3.5-$\sigma$ threshold (vertical dashed black line) -- adopted as the reliability limit for our analysis -- our catalogue is $\sim $\,99\% complete and has a false detection rate of $\sim $\,1.6\%.}
\label{fig:recov}
\end{figure}

Our goal is to construct a catalogue which is deep, i.e.\ includes as many real faint sources as possible, but has a very low spurious source fraction. To determine the search parameters necessary to achieve this we use a suite of simulated ALMA maps.  First we prepare cleaned map which are our actual ALMA maps with all sources $>$\,2.5$\sigma$ identified by our automatic source finder removed. Within the primary beam area of each cleaned map we insert five sources  separated from each other by at least two synthesized beam widths. The inserted sources  cover a wide range in significance ($\sim$\,2--20\,$\sigma$) and follow a steeply declining flux density distribution. We then run our code and extract all sources in the simulated maps, in exactly the same fashion as the science maps. We then determine the recovery rate of our inputs sources and how many spurious sources, not associated with any of our model sources\footnote{Below our initial {\textsc{clean}} threshold our simulated sources are inserted after convolution with the dirty beam instead of the {\textsc{clean}} beam used for brighter sources. The dirty beam side lobes can then be boosted by noise peaks to appear as spurious sources above our detection limit.}, are found. This process is repeated 16 times per map.  Thus, in total we insert 7,400 sources into the 88 ALMA maps.  

Figure \ref{fig:recov} shows the results of our simulations for  the fraction of recovered sources and the fraction of detected sources which are spurious, both as a function of measured peak signal-to-noise ratio.  
The cumulative distribution shows that the source extraction recovers $\sim $\,99\% of all  $>$\,3.5$\sigma$ sources,   while a sample selected above this threshold contains only $\sim $\,1.6\% spurious detections. We therefore adopt a 3.5-$\sigma$ detection limit  for the sample used in our analysis. With respect to the source extraction efficiency we expect that this sample will contain less than two spurious sources and will fail to include only one intrinsically $>$\,3.5-$\sigma$ source.  We also employ the recovered and spurious fractions as a function of detection significance to correct our measured counts in our analysis below.  These corrections become significant only  below the 3.5-$\sigma$ catalogue limit and we  highlight the flux regime most affected by those corrections when discussing the counts.
 
As a final test we compare the total fluxes of the SMGs detected by ALMA to the deboosted fluxes measured for the submillimetre sources in W09. To achieve this we sum the primary-beam-corrected ALMA fluxes of all sources above a given detection significance, within the ALMA primary beam area, weighted by the LABOCA beam to calculate the total flux that would have been seen by LABOCA at the submillimetre source positions from W09. Since the primary beams of both instruments are very similar SMGs detected within the ALMA primary beam area contribute to the flux of a given LABOCA source. To our 3.5-$\sigma$ significance threshold the resulting median ALMA/LABOCA flux density ratio and bootstrap error is 0.83$^{+0.09}_{-0.04}$. If the LABOCA flux scale and the flux deboosting of the LESS sources are accurate, this suggests that a contribution from additional fainter SMGs is required to recover the total flux density within the LABOCA beam.  Integrating the flux in sources down to a 3-$\sigma$ significance limit results in a median flux ratio of 0.97$^{+0.07}_{-0.04}$, consistent with unity. Given the increasing number of spurious detections at lower detection significance we will, however, retain a 3.5-$\sigma$ limit for our analysis 
and discuss the comparison of flux scales in more detail in H13.  

In total we detect 99 individual SMGs at $>$\,3.5\,$\sigma$ in the 88 maps used.  Of these maps, 69 show at least one SMG, 19 maps exhibit two SMGs and four maps have three SMGs within the primary beam.  Hence, 19 ALMA maps ($\sim$\,22\%) do not contain a $>$\,3.5\,$\sigma$ source. The associated LABOCA submillimetre sources have a median deboosted flux of 4.5\,mJy and a median detection significance of 3.75\,$\sigma$, making these sources amongst the faintest in the LESS survey.  The resulting full ALMA SMG catalogue and the exploitation of this catalogue and maps to investigate the multi-wavelength properties of the SMGs will be presented in upcoming publications (H13; Simpson et al.\ in prep.).

\section{Results and discussion}
\label{sec:resdis}
%
%
\begin{figure*}
  \centerline{
  \includegraphics[trim=0.0cm 0.0cm 0.0cm 0.25cm, clip, angle=0,width=\textwidth]{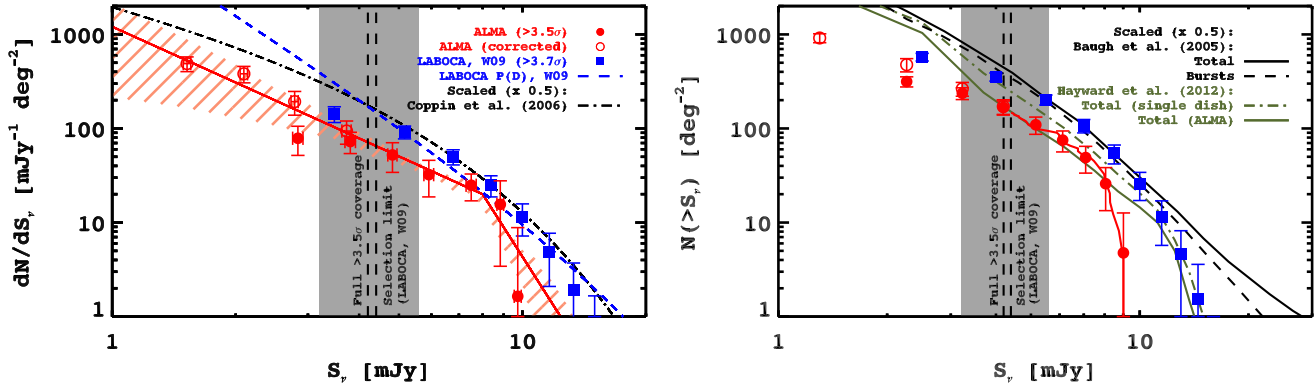}
    }
\caption{{\bf{Left:}} The primary-beam corrected differential 870-$\mu$m source counts from our ALMA survey.
Above the LESS survey limit (depicted with a shaded margin representing the LABOCA rms) the counts of ALMA SMGs should be a true representation of the actual population and any SMG that contributed to the LABOCA source will have been included in our analysis. We also plot counts below this limit for our robust ALMA sample (all bins depicted by filled circles comprise $>$\,3.5-$\sigma$ detections) which is largely uncontaminated by spurious sources while showing a high detection efficiency (see Fig.~\ref{fig:recov}). In addition we extend the counts to $>$\,2.5\,$\sigma$ significance applying the larger corrections necessary for the fractions of spurious and undetected sources.
Fainter than $\sim$\,9\,mJy our data are slightly lower but in reasonable agreement -- within the error margins -- with the scaled fit (scaling adopted from W09) to the number counts from the SHADES survey \citep{COPP06}. However, brighter than $\sim$\,9\,mJy we find a steeper decline than shown in the single-dish counts (a \citet{SCHE76} parameterization of the differential counts resulting from a $P(D)$ analysis of the LESS map by W09 similarly over-predicts the bright counts). Submillimetre sources brighter than this limit were detected by the LESS survey (deboosted values) but no comparably bright SMG is observed by ALMA.  Instead the ALMA maps of the brightest LESS sources  show an increase in source multiplicity.  To parameterise our counts we fit a broken power-law to the data above the LESS survey limit (see Table \ref{tab:fits}), shown by a red solid line and light-red shaded error margins.   {\bf{Right:}} Corresponding cumulative counts of $>$\,3.5-$\sigma$ ALMA sources compared to the deboosted LESS results and the \citet{BAUG05} model predictions. The latter are constrained by observed data from single-dish submillimetre surveys and the predicted fraction of bursts is shown in addition to the total counts. Scaled to the LESS data the model is comparable to our results above the selection limit while the clear mismatch at the bright end due to the absence of corresponding ALMA SMGs is apparent. The scaled \citet{HAYW12} model, designed to predict the ALMA counts, accounts reasonably for source blending effects but also fails to reproduce the bright end. The cumulative counts are corrected for the fraction of missing/spurious detections, but this has a minor effect to our sample due to its high cumulative detection efficiency.}
\label{fig:counts}
\end{figure*}

The ALMA SMGs we have identified can be used to estimate the 870-$\mu$m source counts free from the influence of blending. In the following we describe the derivation of these counts and compare the results to those from previous single-dish surveys.

\subsection{Derivation of ALMA 870~$\mu$m source counts}
\label{sec:rescount}

%
%
\begin{table}
\begin{center}
\caption{Differential and cumulative counts of 870-$\mu$m ALMA-detected SMGs}
\begin{tabular}{llll}
\hline
\hline
 & Differential & & Cumulative\\
$\langle ~S_{\nu} \rangle$ & $dN/dS_{\nu}$ & $~S_{\nu}$ & $N(>S_{\nu})$ \\
$[$mJy$]$ & [mJy$^{-1}$deg$^{-2}$] & [mJy] & [deg$^{-2}$]\\
\hline
$4.8~(0.1)$ & $52.3~(18.2)$ & $4.2$ & 167.3 $~(28.6)$ \\
$5.9~(0.2)$ & $32.3~(13.6)$ & $5.2$ & 109.4 $~(22.8)$ \\
$7.5~(0.2)$ & $24.9~(7.9)$ & $6.1$ & 75.2 $~(18.8)$ \\
$8.8~(0.2)$ & $15.6~(12.2)$ & $7.1$ & 49.1 $~(15.5)$ \\
$9.7~(0.2)$ & $1.6~(7.2)$ & $8.0$ & 25.9 $~(12.5)$ \\
$11.0$ & $0.0$ & $9.0$ & 4.8 $~(7.9)$ \\
$14.0$ & $0.0$ &  &  \\
\hline
\label{tab:counts}
\end{tabular}
\end{center}
\noindent{\footnotesize Notes: Differential (left) and cumulative (right) 870-$\mu$m SMG counts down to the LESS survey
catalogue selection limit as shown in Figure \ref{fig:counts}. The counts throughout the entire flux regime do not require any correcting assumptions. The 1\;$\sigma$ uncertainty ranges stated in brackets are based on the corresponding standard deviations obtained in 1,000 Monte Carlo realizations of our catalog according to the flux uncertainties of individual sources (see \S \ref{sec:rescount} for details). The Poissonian uncertainty is additionally taken into account. 
\hfill }
\end{table}

We are interested in the differential number counts of SMGs as a function of flux. Our ALMA survey covers submillimetre sources above the flux limit of the LESS discovery survey catalogue (W09) so that the effective ALMA survey area, $A_{\rm{ALMA}}$, above this selection limit is given by 

\for{\label{gl:area} A_{\rm{ALMA}} = \frac{N_{\rm{maps}}({\rm ALMA})}{N_{\rm{sources}}(\rm{W09})} \times A_{\rm{LESS}},} 
where $N_{\rm{maps}}({\rm ALMA})=$\,88 is the number of ALMA maps used in this study, $A_{\rm{LESS}}=$\,0.35\,deg$^2$ is the LESS survey area and $N_{\rm{sources}}(\rm{W09})=$\,126 is the total number of submillimetre sources in the LESS catalogue.

Necessarily, this area is only correct in the flux regime covered by the LESS survey.\footnote{Strictly speaking, this  area is only valid for sources in the phase center of a given ALMA map since the sensitivity monotonically drops to 50\% at the primary beam radius. Nevertheless, for the interesting regime of our analysis -- above the LESS survey flux limit -- even at half the phase center sensitivity sources would be detected at a $>$\,3.5\,$\sigma$ level and hence included in our analysis even if residing right at the edge of our field of view.}
Below the LESS flux limit the source counts derived from our ALMA maps must be considered biased since the observations were typically taken in the vicinity of a brighter submillimetre source and so are not necessarily representative of the fainter source population. This restriction already applies  to the faint-end of our sample of $>$\,3.5-$\sigma$ ALMA SMGs (which are below the LESS catalogue flux limit) and we neither attempt any interpretation of these values nor apply further assumptions.  For completeness and purely informational value, we additionally derive differential counts for all $>$\,2.5-$\sigma$ ALMA SMGs.

For the computation of all counts presented here we take into account the flux uncertainty by randomly assigning fluxes to all sources based on their individual error margins. We thereby assume that the individual error distributions are Gaussian and derive the counts for 1,000 resamples. Our best estimate for a given count is given by the mean of all resamples while the standard deviation of which is added in quadrature to the Poissonian error to derive an uncertainty range.

Given our findings in \S\ref{sec:sim}, the source sample and hence the counts are affected by our signal-to-noise ratio (SNR) selection and so we use the recovery and spurious fraction rates derived from our simulations to correct our observed counts. However, we stress that these extraction biases barely affect our sample and even less so the flux regime above the LESS detection limit. Our parameterizations (from Figure~\ref{fig:recov}) of the fractions of spurious detections, $f_{\rm{spurious}}(\rm{SNR})$, and sources recovered in our simulations, $f_{\rm{recovered}}(\rm{SNR})$, provide us with the probability that an SMG with a given SNR is spurious and also the likelihood that  SMGs of the same SNR are missed. For each SMG we therefore derive an individual source probability of $p(\rm{SNR})=2-f_{\rm{recovered}}(\rm{SNR})-f_{\rm{spurious}}(\rm{SNR})$. Corrected differential source counts for a given resample are then given by the total of all source probabilities in a given flux bin and normalized using Equation~(\ref{gl:area}). 

Figure~\ref{fig:counts} shows the resulting differential as well as the cumulative number counts (summarized in Table \ref{tab:counts}) along with the corresponding corrected values. The differential source counts derived for our sample ($>$\,3.5\,$\sigma$) are, as expected, only affected by the bias-corrections at the lowest flux densities while the cumulative counts are even less affected. For flux density bins containing $<$\,3.5-$\sigma$ sources -- far below the LESS survey limit -- we show only bias-corrected values. We also show the best-fit broken power-law to describe our differential counts above the LESS catalogue limit and summarize all these fit parameters in Table~\ref{tab:fits}. The uncertainty range for each parameter is thereby derived by bootstrapping over the parametric fits to all our resamples. 

\subsection{The absence of the bright SMG population}

The most surprising result of our counts is a clear break at the bright end caused by a lack of bright SMGs in our ALMA maps. None of our maps detect an SMG with a flux $\gsim $\,9\,mJy despite twelve sources with 870-$\mu$m fluxes $>$\,9\,mJy in the LESS survey. In at least four of these cases, the submillimetre sources comprise multiple (fainter) SMGs. This is particularly clear in the very brightest LESS sources ($\gsim $\,12\,mJy; see Figure~\ref{fig:maps}) where we detect multiple high-significance ($\gg $\,6$\sigma$) SMGs in each map. In the remaining cases, a single ALMA SMG is detected above 3.5\,$\sigma$, although in all cases, the flux density of this SMG significantly under-predicts the LABOCA flux. This shortfall in flux could arise either from our resolving-out extended emission or from the presence of several  $\sim$\,1--1.5\,mJy SMGs, which lie below our detection threshold.  Nevertheless, an important result of our survey is that the brightest submillimetre sources ($> $\,9\,mJy) in single-dish surveys likely comprise multiple, fainter SMGs. 

On average the mutual separations between different SMGs found in a given map are $>6''$, corresponding to physical scales of $\gg 40$\,kpc in the typical redshift range and provided that we are dealing with physical and not just projected pairs. The very closest projected distance between two SMGs is $2.6''$ as found in only a single map (see Fig. \ref{fig:maps}), corresponding to at least $\sim 20$\,kpc of separation if both SMGs reside at the same redshift. Typically, we find that the flux in a map showing multiple detections is distributed in ratios of 70:30 (double detections) and 50:30:20 (triple detections). Given such clear angular separations we indeed need to count each ALMA component as individual SMGs. We note that the scale, orientation and flux ratio of the multiple components are not consistent configurations caused by gravitational lensing and hence unlikely to represent multiply-imaged sources. 

SMGs with 870-$\mu$m fluxes of $> $\,9\,mJy are likely to be HLIRGs \citep{MRR00,MRR10} if they lie at $z>$\,1. The lack of large numbers of such bright SMGs in our sample  then implies  a natural limit to the star-formation rate in an SMG of $\lsim $\,10$^3$\,M$_\odot$\,yr$^{-1}$ (for a \citet{SALP55} IMF).  This maximal star-formation rate (SFR) is driven by the ratio of the mass of the available gas reservoir and the free-fall time of the system \citep{LEH96}. For a free-fall time of $\sim$\,50\,Myrs as found by \citet{KENN98} for local starbursts, the implied cold gas mass limit is $\lsim $\,$5 \times 10^{10}$\,M$_\odot$.  This is comparable to the limiting gas mass for SMGs found by \citet{BOTH12}. Integrating the parameterisation of our differential counts under consideration of the error margins the absence of very bright SMGs in the LESS survey area therefore suggests that high-redshift galaxies with cold gas masses significantly above $5 \times 10^{10}$\,M$_\odot$ have space densities of $< 10^{-5}$\,Mpc$^{-3}$. For normal star forming $z<3$ galaxies, \citet{KARI11} suggest that their inverse free-fall time constitutes a potential upper limit to their specific SFR. Assuming a typical stellar mass of $\sim 10^{11}$\,M$_\odot$ for our SMGs \citep[e.g.,][]{SWIN12} we also find that their specific SFR is in agreement with this upper limit.  

%
%
\begin{table}
\begin{center}
\caption{Parameterized fit to the differential counts of 870-$\mu$m ALMA-detected SMGs}
\begin{tabular}{lcccc}
\hline
\hline
$dN/dS_{\nu}$ & $N^{\ast}$  & $S^{\ast}_{\nu}$ & $\alpha$ & $\beta$  \\
$[1/N^{\ast} ]$ & [mJy$^{-1}$deg$^{-2}$] & [mJy] & &\\
\hline
$\left(\frac{S_{\nu}}{S^{\ast}_{\nu}} \right)^{-\alpha}, \; S_{\nu} <  S^{\ast}_{\nu}$  &  &  &  & \\
$\left(\frac{S_{\nu}}{S^{\ast}_{\nu}} \right)^{-\beta}, \; S_{\nu} \ge  S^{\ast}_{\nu} $  & $1208^{+902}_{-993}$ & 8 & $2.0^{+0.5}_{-0.4}$ & $6.9^{+2.8}_{-2.3}$ \\
\hline
\label{tab:fits}
\end{tabular}
\end{center}
\noindent{\footnotesize Notes: Best-fit broken power law parameters to describe our data.
The fit is applied only to the data above the LESS catalogue limit and provides a good representation of the steepness of the bright end of our SMG counts (above the fixed breaking point of 8\,mJy). We urge caution when extrapolating this parameterization to much fainter fluxes. 
\hfill }
\end{table}

We therefore parameterise the differential counts with a double power-law with a break point.  Since our data coverage of the regime brighter than the LESS survey limit is really too sparse for a four parameter model fit, we choose to fix the break point at 8\,mJy. Fitting this model we find a factor of $>$\,3 difference in the power law indices of the two power law components (see Table~\ref{tab:fits}). A simple step function would also provide a reasonable representation of our data while not significantly changing the result.

Although previous interferometric surveys of submillimetre sources cannot be considered complete as they comprise a complex mix of follow-up observations from heterogeneous surveys and small survey fields, it is instructive to compare our findings to these previous results.  For example, \citet{BARG12} obtained deep integrations of four $\gsim$\,10\,mJy SMGs in GOODS-N with the SMA at 860\,$\mu$m and showed that at least one breaks up into multiple components across $\sim$\,5$''$. Earlier results by \citet{WANG11} also  suggested that potentially $\sim $\,30\% of $>$\,5-mJy 850-$\mu$m sources could comprise such multiple systems, potentially rising to $>$\,90\% above $\sim$\,8\,mJy \citep[see also][]{WANG07,YOUN08A, COWI09, SMOL12B}.

These results support our interpretation that the number of bright submillimetre sources $\gsim $\,9\,mJy from single-dish surveys have been substantially over-estimated, producing artificially high SMG number counts at the brightest fluxes. To highlight this we  show in Figure~\ref{fig:counts} the results from W09 and those from the SCUBA HAlf Degree Extragalactic Survey (SHADES; \citealp{COPP06}) scaled to the LESS data as described by W09. We refer to both studies for an extensive comparison to other $\sim $\,850-$\mu$m single-dish surveys.  It is noteworthy that even sophisticated methods to estimate the differential source  
counts from single-dish submillimetre surveys -- such as the probabilistic $P(D)$ analysis
presented by W09 -- still recover a false excess of bright sources
(see Figure~\ref{fig:counts}). In order to derive the true counts, this $P(D)$ analysis, 
which derives the differential counts solely from the flux distribution in the map and not
based on individually extracted sources, would need to account for the clustering of 
sources on small angular scales. A similar bias may also be present at intermediate flux 
levels, where our counts are not significantly lower than the single-dish results, as  
would be necessary if source numbers were conserved above our flux limit. This may be 
hinting that these fainter sources also suffer from multiplicity effects. Moreover, some
of the ALMA maps of fainter LABOCA sources which lack $>$\,3.5-$\sigma$ SMGs may then be
explained by the presence of multiple SMGs below our detection limit.\footnote{Assuming
standard dust properties a galaxy at $z>$\,1 with a star-formation rate of  
100\,M$_{\odot}$\,yr$^{-1}$  would have an  870-$\mu$m flux of $\sim $\,1\,mJy, below our
detection limit. For a detailed discussion we refer the reader to H13.} Hence, the statistical prediction by W09 of only five LESS sources being spurious may still be valid.

The high multiplicity of the brightest submillimetre sources also has wider implications. For example, it has been claimed that the brightest (most luminous) SMGs evolve more strongly than fainter systems and hence are preferentially found at the highest redshifts \citep[e.g.,][]{IVIS02, IVIS07, JWALL08, MARS11, SMOL12B}. This could now simply be explained by two effects: confusion, which nullifies the statistical techniques used to identify counterparts, meaning that the very brightest submillimetre sources lack obvious counterparts at other wavelengths and as a result are associated with (undetectable) high-redshift sources; and the significant overestimation of the submillimetre fluxes for the identified counterparts, whose artificially enhanced radio/submillimetre flux ratios then mimic those expected for high-redshift SMGs.

Similarly, the detection rate of $^{12}$CO emission in the SMG survey of \citet{BOTH12} declines with 850-$\mu$m flux.  For example, from their sample of 40 SMGs, the eight of which are undetected (including 2\,/\,5 with $S_{850}\geq $\,10\,mJy) in $^{12}$CO have a median $S_{850}=$\,8.1\,$\pm $\,0.7\,mJy, compared to $S_{850}=$\,5.9\,$\pm$\,0.7\,mJy for the $^{12}$CO detections (this is significant at 92\% confidence level). This modest difference would be explained if a higher fraction of the brighter sources have their submillimetre fluxes boosted by emission from other sources projected along the line of sight within the beam. This scenario is consistent with the suggestion of \citet{WANG11}  that where several SMGs comprise a submillimetre source, those individual components are not necessarily physically associated.  

The high multiplicity of bright submillimetre sources may also bias the form of the far-infrared--radio correlation, which is widely used to infer star formation rates out to high redshifts. For example, where the submillimetre flux of a source is derived from single-dish photometry, and in fact represents contributions from several SMGs, with the radio emission coming from a single source, then this will produce a systematic offset and scatter in the derived far-infrared--radio correlation.  A number of studies \citep[e.g.,][]{IBAR08, SARG10A, IVIS10, BOUR11} have attempted to trace the evolution of this relation over a range of redshifts, of which several  focussed on submillimetre-bright sources \citep[e.g.,][]{KOVA06, MURP09, MURP09B}. The latter reported  a radio excess for submillimetre-selected sources compared to local star-forming galaxies and -- if true -- this offset would increase further if only a single sub-component is associated with the radio emitter.\footnote{It should be noted  that \citet{BARG12} find that five of their SMA sources  agree with the local relation.} However, we postpone a discussion of the radio properties of our ALMA SMGs to an upcoming publication.  

Finally, we also compare our SMG counts in Figure~\ref{fig:counts} to the predicted counts from the model of \citet{BAUG05} which, by design, match the single-dish observations. In comparison, consequently, our counts are significantly lower, particularly at the bright end. Since \citet{BAUG05} could only reproduce the submillimetre counts by including a top-heavy IMF during bursts of star formation, the tension created by our ALMA counts might aid the model to reproduce the true counts without resorting to a non-standard IMF. However, our data show no strong disagreement at intermediate fluxes so that the fundamental problem for models of reproducing the normalisation of the counts may still persist.\footnote{E.g., \citet{FONT07} highlight the difficulty of semi-analytical models to simultaneously reproduce the abundance of SMGs and $z \lsim 1$ massive galaxies when a standard IMF is used.}
Recently, \citet{HAYW12} predicted the submillimetre source counts by a combined semi-analytical/hydrodynamic approach without including a top-heavy IMF for bursts. Throughout the flux regime covered by our study, they predict a consistently high fraction ($> $\,30\%) of galaxy pairs, in agreement with our observations. Still, their predicted bright SMG counts do not drop as sharply as our data indicate since our fraction of multiple sources steeply rises at the bright end\footnote{For a detailed analysis of the source multiplicity as a function of flux we refer the
reader to H13. Note that the interferometric follow-up of LABOCA sources in the COSMOS
field presented by \citet{SMOL12B} shows a similarly elevated fraction of multiple sources
at the bright end.}. At fluxes below the break and the above the LESS survey limit, Figure \ref{fig:counts} shows that their model reproduces the marginal difference between the counts at interferometric and single dish resolution as a result of blended, physically separated not interacting galaxy pairs, in agreement with our study. It should be noted, however, that in contrast to, e.g., the \citet{BAUG05} model the predictive power of the \citet{HAYW12} work is limited to the submillimetre properties of distant star forming galaxies, as it is not required to reproduce the global population of galaxies, particularly in the local Universe.

\section{Summary} 

We have presented source number counts derived from an 870-$\mu$m ALMA survey of submillimetre sources from the 870-$\mu$m LABOCA survey of the Extended {\it Chandra} Deep Field South by W09. Compared to the parent survey, our ALMA maps are three times deeper and have an angular resolution which is  an order of magnitude higher, allowing us to remove the influence of blending on the counts above the LABOCA detection limit. We find that our source counts are in broad agreement with those of the LABOCA survey and previous literature results.  However, brighter than $\sim$\,8\,mJy our counts show a deficit of sources compared to those from single-dish surveys. This is caused by multiple SMGs, which are found to be well separated (typically by $\sim 6''$) at
the $\sim$\,1.5$''$ resolution of our ALMA maps, being blended into single sources at the resolution of the single-dish surveys.  This trend has also been seen in recent studies of smaller samples of submillimetre sources.
Our results suggest that multiplicity in submillimetre sources is significant at the brightest fluxes, but may also influence fainter submillimetre sources, $\sim$\,4\,mJy, from single-dish surveys.  The absence of bright SMGs in our sample  
implies  a  limit to the maximum star-formation rate in an SMG of $\lsim $\,10$^3$\,M$_\odot$\,yr$^{-1}$ (for a Salpeter IMF), which in turn suggests that systems with  gas masses in excess of $\sim $\,$5 \times10^{10}$\,M$_\odot$ have space densities of $< 10^{-5}$\,Mpc$^{-3}$ at $z \gsim 1$.

\section*{Acknowledgments}
We thank T.\ Muxlow, G.\ Bendo and the team of the Manchester ALMA RC node for their support as well as Amy Barger for helpful comments on the manuscript. 
We thank the anonymous referee for providing very helpful suggestions to improve the quality of this paper.
AK acknowledges support from STFC and AMS and TRG gratefully acknowledge STFC Advanced Fellowships. IRS acknowledges support from STFC and a Leverhume Fellowship. KEKC acknowledges support from the endowment of the Lorne Trottier Chair in Astrophysics and Cosmology at McGill, the Natural Science and Engineering Research Council of Canada (NSERC), and a L'Or\'{e}al Canada for Women in Science Research Excellence Fellowship, with the support of the Canadian Commission for UNESCO. KK thanks the Swedish Research Council for support. The ALMA observations were carried out under program ADS/JAO.ALMA\#2011.0.00294.S. ALMA is a partnership of ESO (representing its member states), NSF (USA) and NINS (Japan), together with NRC (Canada) and NSC and ASIAA (Taiwan), in cooperation with the Republic of Chile. The Joint ALMA Observatory is operated by ESO, AUI/NRAO and NAOJ. This publication is based on data acquired with the APEX under programme IDs 078.F-9028(A), 079.F-9500(A), 080.A-3023(A) and 081.F-9500(A). APEX is a collaboration between the Max-Planck-Institut f\"ur Radioastronomie, the European Southern Observatory and the Onsala Space Observatory.
\bibliographystyle{mn2e.bst} 
\bibliography{bibliography_ak.bib}

\begin{thebibliography}{}

\bibitem[\protect\citeauthoryear{{Alexander} et~al.,}{{Alexander}
  et~al.}{2008}]{ALEX08}
{Alexander} D.~M.  et~al., 2008, \apj, 687, 835

\bibitem[\protect\citeauthoryear{{Barger}, {Cowie}, {Sanders}, {Fulton},
  {Taniguchi}, {Sato}, {Kawara} \& {Okuda}}{{Barger} et~al.}{1998}]{BARG98}
{Barger} A.~J.,  {Cowie} L.~L.,  {Sanders} D.~B.,  {Fulton} E.,  {Taniguchi}
  Y.,  {Sato} Y.,  {Kawara} K.,    {Okuda} H.,  1998, \nat, 394, 248

\bibitem[\protect\citeauthoryear{{Barger}, {Wang}, {Cowie}, {Owen}, {Chen} \&
  {Williams}}{{Barger} et~al.}{2012}]{BARG12}
{Barger} A.~J.,  {Wang} W.-H.,  {Cowie} L.~L.,  {Owen} F.~N.,  {Chen} C.-C.,
  {Williams} J.~P.,  2012, \apj, 761, 89

\bibitem[\protect\citeauthoryear{{Baugh}, {Lacey}, {Frenk}, {Granato}, {Silva},
  {Bressan}, {Benson} \& {Cole}}{{Baugh} et~al.}{2005}]{BAUG05}
{Baugh} C.~M.,  {Lacey} C.~G.,  {Frenk} C.~S.,  {Granato} G.~L.,  {Silva} L.,
  {Bressan} A.,  {Benson} A.~J.,    {Cole} S.,  2005, \mnras, 356, 1191

\bibitem[\protect\citeauthoryear{{Blain}, {Kneib}, {Ivison} \& {Smail}}{{Blain}
  et~al.}{1999}]{BLAI99}
{Blain} A.~W.,  {Kneib} J.-P.,  {Ivison} R.~J.,    {Smail} I.,  1999, \apjl,
  512, L87

\bibitem[\protect\citeauthoryear{{Bothwell} et~al.,}{{Bothwell}
  et~al.}{2012}]{BOTH12}
{Bothwell} M.~S.  et~al., 2012, MNRAS, in press, arXiv:astro-ph/1205.1511

\bibitem[\protect\citeauthoryear{{Bourne}, {Dunne}, {Ivison}, {Maddox},
  {Dickinson} \& {Frayer}}{{Bourne} et~al.}{2011}]{BOUR11}
{Bourne} N.,  {Dunne} L.,  {Ivison} R.~J.,  {Maddox} S.~J.,  {Dickinson} M.,
  {Frayer} D.~T.,  2011, \mnras, 410, 1155

\bibitem[\protect\citeauthoryear{{Chapman}, {Blain}, {Smail} \&
  {Ivison}}{{Chapman} et~al.}{2005}]{CHAP05}
{Chapman} S.~C.,  {Blain} A.~W.,  {Smail} I.,    {Ivison} R.~J.,  2005, \apj,
  622, 772

\bibitem[\protect\citeauthoryear{{Condon}}{{Condon}}{1997}]{COND97}
{Condon} J.~J.,  1997, \pasp, 109, 166

\bibitem[\protect\citeauthoryear{{Coppin} et~al.,}{{Coppin}
  et~al.}{2006}]{COPP06}
{Coppin} K.  et~al., 2006, \mnras, 372, 1621

\bibitem[\protect\citeauthoryear{{Cowie}, {Barger}, {Wang} \&
  {Williams}}{{Cowie} et~al.}{2009}]{COWI09}
{Cowie} L.~L.,  {Barger} A.~J.,  {Wang} W.-H.,    {Williams} J.~P.,  2009,
  \apjl, 697, L122

\bibitem[\protect\citeauthoryear{{Dannerbauer}, {Lehnert}, {Lutz}, {Tacconi},
  {Bertoldi}, {Carilli}, {Genzel} \& {Menten}}{{Dannerbauer}
  et~al.}{2002}]{DANN02}
{Dannerbauer} H.,  {Lehnert} M.~D.,  {Lutz} D.,  {Tacconi} L.,  {Bertoldi} F.,
  {Carilli} C.,  {Genzel} R.,    {Menten} K.,  2002, \apj, 573, 473

\bibitem[\protect\citeauthoryear{{Fontanot}, {Monaco}, {Silva} \&
  {Grazian}}{{Fontanot} et~al.}{2007}]{FONT07}
{Fontanot} F.,  {Monaco} P.,  {Silva} L.,    {Grazian} A.,  2007, \mnras, 382,
  903

\bibitem[\protect\citeauthoryear{{Gear}, {Lilly}, {Stevens}, {Clements},
  {Webb}, {Eales} \& {Dunne}}{{Gear} et~al.}{2000}]{GEAR00}
{Gear} W.~K.,  {Lilly} S.~J.,  {Stevens} J.~A.,  {Clements} D.~L.,  {Webb}
  T.~M.,  {Eales} S.~A.,    {Dunne} L.,  2000, \mnras, 316, L51

\bibitem[\protect\citeauthoryear{{Giavalisco} et~al.,}{{Giavalisco}
  et~al.}{2004}]{GIAV04}
{Giavalisco} M.  et~al., 2004, \apjl, 600, L93

\bibitem[\protect\citeauthoryear{{Granato}, {Silva}, {Lapi}, {Shankar}, {De
  Zotti} \& {Danese}}{{Granato} et~al.}{2006}]{GRAN06}
{Granato} G.~L.,  {Silva} L.,  {Lapi} A.,  {Shankar} F.,  {De Zotti} G.,
  {Danese} L.,  2006, \mnras, 368, L72

\bibitem[\protect\citeauthoryear{{Hayward}, {Narayanan}, {Kere{\v s}},
  {Jonsson}, {Hopkins}, {Cox} \& {Hernquist}}{{Hayward} et~al.}{2012}]{HAYW12}
{Hayward} C.~C.,  {Narayanan} D.,  {Kere{\v s}} D.,  {Jonsson} P.,  {Hopkins}
  P.~F.,  {Cox} T.~J.,    {Hernquist} L.,  2012, MNRAS, in press,
  arXiv:astro-ph/1209.2413

\bibitem[\protect\citeauthoryear{{Hickox} et~al.,}{{Hickox}
  et~al.}{2012}]{HICK12}
{Hickox} R.~C.  et~al., 2012, \mnras, 421, 284

\bibitem[\protect\citeauthoryear{{Hodge} et~al.,}{{Hodge}
  et~al.}{2013}]{HODG13}
{Hodge} J.  et~al., 2013, ApJ, submitted

\bibitem[\protect\citeauthoryear{{Hopkins}, {Afonso}, {Chan}, {Cram},
  {Georgakakis} \& {Mobasher}}{{Hopkins} et~al.}{2003}]{HOPK03}
{Hopkins} A.~M.,  {Afonso} J.,  {Chan} B.,  {Cram} L.~E.,  {Georgakakis} A.,
  {Mobasher} B.,  2003, \aj, 125, 465

\bibitem[\protect\citeauthoryear{{Hughes} et~al.,}{{Hughes}
  et~al.}{1998}]{HUGH98}
{Hughes} D.~H.  et~al., 1998, \nat, 394, 241

\bibitem[\protect\citeauthoryear{{Ibar} et~al.,}{{Ibar} et~al.}{2008}]{IBAR08}
{Ibar} E.  et~al., 2008, \mnras, 386, 953

\bibitem[\protect\citeauthoryear{{Ivison} et~al.,}{{Ivison}
  et~al.}{2007}]{IVIS07}
{Ivison} R.~J.  et~al., 2007, \mnras, 380, 199

\bibitem[\protect\citeauthoryear{{Ivison} et~al.,}{{Ivison}
  et~al.}{2002}]{IVIS02}
{Ivison} R.~J.  et~al., 2002, \mnras, 337, 1

\bibitem[\protect\citeauthoryear{{Ivison} et~al.,}{{Ivison}
  et~al.}{2010}]{IVIS10}
{Ivison} R.~J.  et~al., 2010, \aap, 518, L35

\bibitem[\protect\citeauthoryear{{Karim} et~al.,}{{Karim}
  et~al.}{2011}]{KARI11}
{Karim} A.  et~al., 2011, \apj, 730, 61

\bibitem[\protect\citeauthoryear{{Kennicutt} Jr.}{{Kennicutt}}{1998}]{KENN98}
{Kennicutt} Jr. R.~C.,  1998, \apj, 498, 541

\bibitem[\protect\citeauthoryear{{Kov{\'a}cs}, {Chapman}, {Dowell}, {Blain},
  {Ivison}, {Smail} \& {Phillips}}{{Kov{\'a}cs} et~al.}{2006}]{KOVA06}
{Kov{\'a}cs} A.,  {Chapman} S.~C.,  {Dowell} C.~D.,  {Blain} A.~W.,  {Ivison}
  R.~J.,  {Smail} I.,    {Phillips} T.~G.,  2006, \apj, 650, 592

\bibitem[\protect\citeauthoryear{{Lehnert} \& {Heckman}}{{Lehnert} \&
  {Heckman}}{1996}]{LEH96}
{Lehnert} M.~D.,  {Heckman} T.~M.,  1996, \apj, 462, 651

\bibitem[\protect\citeauthoryear{{Lutz} et~al.,}{{Lutz} et~al.}{2001}]{LUTZ01}
{Lutz} D.  et~al., 2001, \aap, 378, 70

\bibitem[\protect\citeauthoryear{{Marsden} et~al.,}{{Marsden}
  et~al.}{2011}]{MARS11}
{Marsden} G.  et~al., 2011, \mnras, 417, 1192

\bibitem[\protect\citeauthoryear{{McMullin}, {Waters}, {Schiebel}, {Young} \&
  {Golap}}{{McMullin} et~al.}{2007}]{MCMU07}
{McMullin} J.~P.,  {Waters} B.,  {Schiebel} D.,  {Young} W.,    {Golap} K.,
  2007, in {Shaw} R.~A.,  {Hill} F.,   {Bell} D.~J.,  eds,  Astronomical
  Society of the Pacific Conference Series Vol. 376, Astronomical Data Analysis
  Software and Systems XVI. p.~127

\bibitem[\protect\citeauthoryear{{Murphy}}{{Murphy}}{2009}]{MURP09B}
{Murphy} E.~J.,  2009, \apj, 706, 482

\bibitem[\protect\citeauthoryear{{Murphy}, {Chary}, {Alexander}, {Dickinson},
  {Magnelli}, {Morrison}, {Pope} \& {Teplitz}}{{Murphy} et~al.}{2009}]{MURP09}
{Murphy} E.~J.,  {Chary} R.-R.,  {Alexander} D.~M.,  {Dickinson} M.,
  {Magnelli} B.,  {Morrison} G.,  {Pope} A.,    {Teplitz} H.~I.,  2009, \apj,
  698, 1380

\bibitem[\protect\citeauthoryear{{Rowan-Robinson}}{{Rowan-Robinson}}{2000}]{MR%
R00}
{Rowan-Robinson} M.,  2000, \mnras, 316, 885

\bibitem[\protect\citeauthoryear{{Rowan-Robinson} \& {Wang}}{{Rowan-Robinson}
  \& {Wang}}{2010}]{MRR10}
{Rowan-Robinson} M.,  {Wang} L.,  2010, \mnras, 406, 720

\bibitem[\protect\citeauthoryear{{Salpeter}}{{Salpeter}}{1955}]{SALP55}
{Salpeter} E.~E.,  1955, \apj, 121, 161

\bibitem[\protect\citeauthoryear{{Sargent} et~al.,}{{Sargent}
  et~al.}{2010}]{SARG10A}
{Sargent} M.~T.  et~al., 2010, \apjs, 186, 341

\bibitem[\protect\citeauthoryear{Schechter}{Schechter}{1976}]{SCHE76}
Schechter P.,  1976, \apj, 203, 297

\bibitem[\protect\citeauthoryear{{Schinnerer} et~al.,}{{Schinnerer}
  et~al.}{2004}]{SCHI04}
{Schinnerer} E.  et~al., 2004, \aj, 128, 1974

\bibitem[\protect\citeauthoryear{{Schinnerer} et~al.,}{{Schinnerer}
  et~al.}{2010}]{SCHI10}
{Schinnerer} E.  et~al., 2010, \apjs, 188, 384

\bibitem[\protect\citeauthoryear{{Smail}, {Ivison} \& {Blain}}{{Smail}
  et~al.}{1997}]{SMAI97}
{Smail} I.,  {Ivison} R.~J.,    {Blain} A.~W.,  1997, \apjl, 490, L5

\bibitem[\protect\citeauthoryear{{Smol{\v c}i{\'c}} et~al.,}{{Smol{\v c}i{\'c}}
  et~al.}{2012}]{SMOL12B}
{Smol{\v c}i{\'c}} V.  et~al., 2012, \aap, 548, A4

\bibitem[\protect\citeauthoryear{{Swinbank}, {Chapman}, {Smail}, {Lindner},
  {Borys}, {Blain}, {Ivison} \& {Lewis}}{{Swinbank} et~al.}{2006}]{SWIN06}
{Swinbank} A.~M.,  {Chapman} S.~C.,  {Smail} I.,  {Lindner} C.,  {Borys} C.,
  {Blain} A.~W.,  {Ivison} R.~J.,    {Lewis} G.~F.,  2006, \mnras, 371, 465

\bibitem[\protect\citeauthoryear{{Swinbank} et~al.,}{{Swinbank}
  et~al.}{2012}]{SWIN12}
{Swinbank} A.~M.  et~al., 2012, \mnras, 427, 1066

\bibitem[\protect\citeauthoryear{{Wall}, {Pope} \& {Scott}}{{Wall}
  et~al.}{2008}]{JWALL08}
{Wall} J.~V.,  {Pope} A.,    {Scott} D.,  2008, \mnras, 383, 435

\bibitem[\protect\citeauthoryear{{Wang}, {Cowie} \& {Barger}}{{Wang}
  et~al.}{2004}]{WANG04}
{Wang} W.-H.,  {Cowie} L.~L.,    {Barger} A.~J.,  2004, \apj, 613, 655

\bibitem[\protect\citeauthoryear{{Wang}, {Cowie}, {Barger} \&
  {Williams}}{{Wang} et~al.}{2011}]{WANG11}
{Wang} W.-H.,  {Cowie} L.~L.,  {Barger} A.~J.,    {Williams} J.~P.,  2011,
  \apjl, 726, L18

\bibitem[\protect\citeauthoryear{{Wang}, {Cowie}, {van Saders}, {Barger} \&
  {Williams}}{{Wang} et~al.}{2007}]{WANG07}
{Wang} W.-H.,  {Cowie} L.~L.,  {van Saders} J.,  {Barger} A.~J.,    {Williams}
  J.~P.,  2007, \apjl, 670, L89

\bibitem[\protect\citeauthoryear{{Wei{\ss}} et~al.,}{{Wei{\ss}}
  et~al.}{2009}]{WEIS09}
{Wei{\ss}} A.  et~al., 2009, \apj, 707, 1201

\bibitem[\protect\citeauthoryear{{Windhorst}, {van Heerde} \&
  {Katgert}}{{Windhorst} et~al.}{1984}]{WIND84}
{Windhorst} R.~A.,  {van Heerde} G.~M.,    {Katgert} P.,  1984, \aaps, 58, 1

\bibitem[\protect\citeauthoryear{{Younger} et~al.,}{{Younger}
  et~al.}{2007}]{YOUN07}
{Younger} J.~D.  et~al., 2007, \apj, 671, 1531

\bibitem[\protect\citeauthoryear{{Younger} et~al.,}{{Younger}
  et~al.}{2008a}]{YOUN08A}
{Younger} J.~D.  et~al., 2008a, \mnras, 387, 707

\bibitem[\protect\citeauthoryear{{Younger} et~al.,}{{Younger}
  et~al.}{2008b}]{YOUN08B}
{Younger} J.~D.  et~al., 2008b, \apj, 688, 59

\bibitem[\protect\citeauthoryear{{Younger} et~al.,}{{Younger}
  et~al.}{2009}]{YOUN09}
{Younger} J.~D.  et~al., 2009, \apj, 704, 803

\end{thebibliography}

\label{lastpage}

\end{document}